# Measurement of photon correlations with multipixel photon counters

Dmitry Kalashnikov, and Leonid Krivitsky*

*Data Storage Institute, Agency for Science Technology and Research (A-STAR), 5 Engineering Drive 1, 117608 Singapore*
*Corresponding author: Leonid_Krivitskiy@dsi.a-star.edu.sg*

Development of reliable photon number resolving detectors (PNRD), devices which are capable to distinguish 1,2,3.. photons, is of a great importance for quantum optics and its applications. A new class of affordable PNRD is based on multipixel photon counters (MPPC). Here we review results of experiments on using MPPCs for direct characterization of squeezed vacuum (SV) states, generated via parametric downconversion (PDC). We use MPPCs to measure the second order normalized intensity correlation function ($g^{(2)}$) and directly detect the two-mode squeezing of SV states. We also present a method of calibration of crosstalk probability in MPPCs based on $g^{(2)}$ measurements of coherent states.

## 1. Introduction

Multiphoton entangled states hold the promise to advance both practical and fundamental aspects of future quantum technologies. They play an important role in linear optical quantum computation [1-3], quantum metrology [4], imaging [5, 6], and tests of fundamentals of quantum mechanics [7]. Accurate measurement of such states is also of crucial importance for security analysis in quantum cryptography [8], and in implementation of quantum teleportation [9] and entanglement swapping [10] protocols.

Two major challenges in experimental studies of multiphoton entangled states are their generation and measurement. The first challenge can be addressed, for example, by using pulsed laser with high-peak power values in various nonlinear optical processes, possibly enhanced by using cavities and/or nonlinear waveguides. The detailed analysis of the generation problem is outside the scope of this paper, and we direct readers to detailed reviews on this subject [11, 12]. The measurement problem will be addressed here in greater details. We consider direct detection of multiphoton states by using specialized photon number resolving detectors (PNRD) – devices where the produced outcome corresponds to the number of photons in an optical pulse.

Conventional single photon detectors, such as avalanche photodiode (APD) and photomultiplier tubes (PMT), discriminate only between "zero photons" and "one photon or more" because of the "dead time" effect [13]. The straightforward way to overcome this limitation relies upon splitting the incoming optical pulse into different spatial (or temporal) modes, and detection of single photons in each mode by independent APDs [14-16]. The photon number distribution is then derived from the joint statistics of APD responses. Although a remarkable resolution of up to 9 photons was demonstrated with temporal multiplexing, future expansion of such schemes is limited by the necessity to use more beamsplitters and photodetectors, which makes such schemes bulky and challenging to operate. Moreover, since the detection of multiphoton events requires simultaneous firing of several detectors, their non-unity quantum efficiency leads to the loss of a large fraction of multiphoton events.

The next approach is based on using cryogenic detectors, such as visible-light photon counters (VLPC) and transition edge sensors (TES) [17-22]. These devices have moderate photon number resolution (up to 7-9 photons), low noise (<1 c/s), and high quantum efficiency (~80%). One of the approaches relies upon integration of several niobium nitride superconducting nanowire single photon detectors into a waveguide, which results in low jitter and fast recovery time [23]. However, the main limitation of the mentioned above technologies is the requirement of cryogenic cooling (down to 0.1K for TES, 6K for VLPC, and below 10K for superconducting nanowire), which is expensive and demands highly skilled operation.

Another class of devices can be used as PNRD without the necessity of cryogenic cooling. Here we mention a hybrid photomultiplier tube, which demonstrates discrimination of up to 3 photons [24]. A resolution of up to 5 photons was demonstrated with a single InGAs detector by careful characterization of the early stage of the avalanche [25]. Another approach is based on Si-APD with a strongly modulated lateral electric field profile realized by modulated doping profile of p-n junctions [26]. The high speed and discrimination of up to 4 photons has been demonstrated, but the realized device has comparatively low quantum efficiency (~8 %) and exhibits the space-charge effect, which prevents resolution of higher number of photons.

Recently a new approach for realization of the PNRD was suggested based on silicon multipixel photon counters (MPPCs) [27]. In MPPC several hundred silicon APDs, referred to as pixels, are integrated into a single chip of several millimeter size, see Fig.1a,b. Each pixel outputs a pulse signal when it detects a photon. Signals from all the fired pixels are summed at the output of the MPPC. The chip of the MPPC is illuminated by a diffused light in order to minimize the probability of a single pixel to be hit by several photons. Thus the amplitude of the MPPC output is, in an ideal case, equivalent to the number of detected incident photons, see Fig.1c. The concept of MPPC is in some sense similar to the traditional approach of separating an incoming pulse into multiple spatial modes and detection by independent APDs, which is similar to the concept of single photon CCD cameras (EMCCD, ICCD). However, it provides striking advantages in compactness, ease-of-use, photon-number resolution, and cost. Currently, MPPCs find their applications in quantum optics experiments [28-33], as well as in high-energy physics experiments as photosensors for scintillate detectors [34].

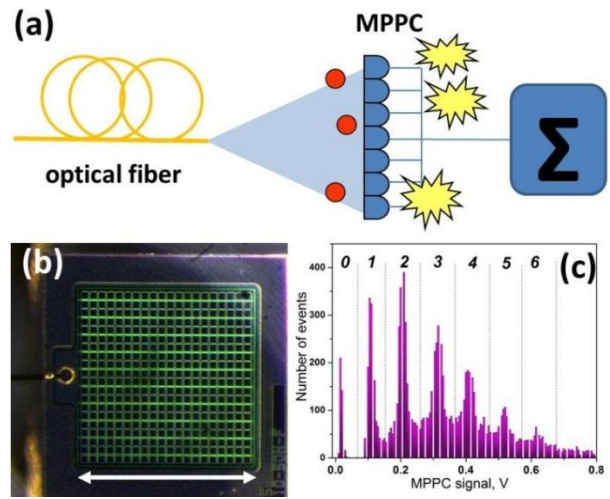

Fig.1 (a) Principle of MPPC operation. A diffused light is impinging on the MPPC matrix, consisting of few hundreds of APD pixels. Signals from all the pixels are summarized at the output. (b) Photo of the MPPC chip with 400 pixels. The scale bar corresponds to 1.5 mm. (c) Example of a histogram of the MPPC output for coherent state. Different peaks correspond to the resolution of photon numbers.

The major drawbacks of MPPC are relatively high dark counts, and the optical crosstalk among pixels. The high dark count rate is due to a large number of individual APDs in the MPPC matrix, with each of the APD gives its contribution to the dark noise, contributing a bit to the dark noise. The dark noise can be significantly reduced by cooling MPPC and providing time gating to the signal. In contrast, the crosstalk requires more elaborated attention. The crosstalk occurs because secondary photons, which are re-emitted during the avalanche in the pixel, may trigger simultaneous additional photon counts in neighboring pixels [35, 36]. Since the crosstalk happens almost simultaneously with the detection event, it is undistinguishable from the actual counts, and thus its impact requires accurate theoretical modelling and characterization.

In this paper we review experiments on using MPPCs for characterization of various entangled multiphoton states. Primary objects of our studies are *squeezed vacuum* (SV) states, generated via parametric downconversion (PDC), which is the nonlinear process producing pairs of strongly correlated photons. We explore different regimes of PDC and measure the normalized second-order intensity correlation function of SV states, and also directly detect the two-mode squeezing. We also present a novel method of calibration of MPPC crosstalk probability based on the measurement of the correlation function of coherent light.

The paper is organized as follows. In Section 2 we derive the algorithm of obtaining the normalized second-order correlation function ($g^{(2)}$) from the MPPC data, introduce a model of the MPPC crosstalk, and model the MPPC saturation. In Section 3 we describe an experiment, where we use the MPPC to measure $g^{(2)}$ of SV states with much less than one photon per pulse. In Section 4 we present a method of calibration of the crosstalk probability of MPPCs based on $g^{(2)}$ measurements of coherent states, and test it with several commercially available devices. In Section 5 we describe an experiment, where MPPCs are used for direct observation of the two-mode squeezing of relatively bright SV states with up to 10 photons per pulse. The results are summarized in the Conclusions section.

## 2. Theoretical modelling of the MPPC

### 2.1 Measurement of correlation functions

Characterization of various states of light in quantum optics is conveniently performed in terms of normalized Glauber's correlation functions (CFs) [37]. The $l-th$ order CF at zero time delay is defined as

$$g^{(l)} = \langle a^{+l} a^l \rangle / \langle a^+ a \rangle^l \quad (1)$$

Where $a^+$, $a$ are photon creation and annihilation operators, respectively. The remarkable feature of normalized CFs is that they are insensitive to optical losses and finite efficiencies of the detectors, which may be not always precisely known in experiments [38]. Here we focus on the analysis of the second-order CF, $l = 2$.

Conventionally $g^{(2)}$ is measured in a Hanbury-Brown and Twiss (HBT) setup consisting of two single photon detectors with their outputs addressed to a coincidence circuit. Let $N_{Coinc}$ be the measured number of coincidences of photocounts of two detectors, and $N_{D1,D2}$ is the measured number of photocounts of individual photodetectors, then $g^{(2)}$ is given by:

$$g^{(2)} \propto \frac{N_{Coinc}}{N_{D1} N_{D2}} \quad (2)$$

Let us now consider measurement of $g^{(2)}$ using the MPPC. We can treat each pair of pixels in the MPPC matrix as a single HBT setup. The number of such HBT setups is equal to the number of 2-combination of $m$: $C_2^m = m(m-1)/2 \approx m^2/2$, where $m$ is the total number of pixels in the matrix (typically $m \geq 100$). Let $N_k$ be the number of detected $k -$ photon events ($k = 1,2,3 ...$). The total number of pairwise coincidences is given by $N_{Coinc} = \sum_{k=2}^{\infty} C_2^k N_k$, where $C_2^k$ is the number of 2-combinations in $k$, which describes the contribution of a detected $k -$photon event into two-fold coincidences. The total amount of photocounts is given by $N_{Total} = \sum_{k=1}^{\infty} k N_k$. In analogy with Eq.(2), $g^{(2)}$ is calculated as a ratio of the number of pairwise coincidences *per one HBT setup* to the squared total number of detected photons *per one detector (pixel)*:

$$g^{(2)} = \left( \frac{N_{Coinc}/(m^2/2)}{N_{Total}/m} \right) = 2 \sum_{k=2}^{\infty} C_2^k N_k / (\sum_{k=1}^{\infty} k N_k)^2 \quad (3)$$

Thus, once the histogram of detected $k$-photon events $N_k$ is obtained in the experiment, see Fig. 1c, $g^{(2)}$ can be directly calculated from Eq.(3). Moreover, any spatially resolving single-photon detector can be used for these measurements, such as, for instance, single-photon sensitive charge coupling devices (EMCCD, ICCD) [39, 40].

### 2.2 Modelling the crosstalk in MPPC

Let us now take into account the MPPC crosstalk, which is introduced through the crosstalk probability for one pixel $p$ The model under consideration is restricted to the second order crosstalk under the assumption of low mean photon number of the detected light, and the following condition for the crosstalk probability has to be fulfilled $p + 2p^2 \gg 3p^3$. One can express the number of $k$-photon events $N_{k,CT}$ with crosstalk probability through the corresponding number of events in the absence of crosstalk $N_k$ as follows:

$$k = 1, N_{1,CT} = N_1 - pN_1 - p^2 N_1, \quad (4a)$$

Where the second term on the right corresponds to the number of single photon events converted to two-photon events, and the third term is the number of single photon events converted to 3-photon events.

$$k = 2, N_{2,CT} = N_2 - 2pN_2 + pN_1 - 2p^2 N_2, \quad (4b)$$

Where the second term on the right corresponds to the number of two-photon events converted into three-photon events, the third term arises due to the impact from one-photon events, and the fourth term describes two-photon events influenced by the double crosstalk. Finally,

$$k > 2, N_{k,CT} = N_k - kpN_k + (k-1)pN_{k-1} + (k-2)p^2 N_{k-2} - kp^2 N_k \quad (4c)$$

In Eq.(4c) the second term on the right is the number of $k$ −photon events being converted by crosstalk into $(k+1)$ −photon events, and the third term is the number of $(k-1)$ −photon events that gained an extra photon due to crosstalk to become $k$ −photon events; the fourth term represents the number of $(k-2)$ −photon events converted into $k$-photon events due to double crosstalk, and the fifth term is the number of $k$ −photon events which contributed to $(k+2)$ −photon events, also due to double crosstalk.

Assuming the crosstalk model mentioned above, the numerator and the denominator of Eq.3 in the presence of the crosstalk take the form:

$$N_{Coinc,CT} = (1 + 2p + 4p^2) \sum_{k=2}^{\infty} C_2^k N_k +$$

$$p(1+3p) \sum_{k=1}^{\infty} k N_k, \quad (5a)$$

$$N_{Total,CT} = (1 + p + 2p^2) \sum_{k=1}^{\infty} k N_k. \quad (5b)$$

From Eqs.(3, 5a, 5b) $g^{(2)}$ takes the following form:

$$g^{(2)}(N_{Total,CT}) = A g_0^{(2)} + B \frac{1}{N_{Total,CT}}, \quad (6)$$

where $g_0^{(2)} = 2 \sum_{k=2}^{\infty} C_2^k N_k / (\sum_{k=1}^{\infty} k N_k)^2$ is the initial second order correlation of light, and

$$A \equiv \frac{1+2p+4p^2}{(1+p+2p^2)^2}, \qquad B \equiv \frac{2p(1+3p)}{1+p+2p^2}.$$

There are two main conclusions followed from Eq.(6). First, the $g^{(2)}$ measured by the MPPC, is different from the $g_0^{(2)}$ of the incident light by a normalization coefficient and an additive term scaling as $1/N_{Total,CT}$. Second, the crosstalk probability $p$ can be easily measured using some light source with known $g_0^{(2)}$, for instance coherent light with $g_0^{(2)} = 1$. Both of these aspects will be addressed experimentally in Sections 3 and 4.

## 2.3 Modelling saturation in MPPC

In case when beams of relatively high intensities are considered, saturation of the MPPCs has to be taken into account by assuming that each MPPC can only resolve up to $N_{max}$ photocounts, given by the number of illuminated pixels. When more than $N_{max}$ photons incident on the detector, not more than $N_{max}$ photocounts will be produced. For modeling of a realistic MPPC, we follow [28, 30], where the loss and crosstalk are considered as Bayesian processes. Let us denote $\langle \hat{n} \rangle$ as the mean number of photons impinging on the detector, $p$ as a crosstalk probability, and $\eta$ as a quantum efficiency of each MPPC pixel, which also accounts for losses in an optical system. The positive-operator valued measure (POVM) effecting the measurement of $m$ photons with crosstalk with one MPPC, for $m = 0, 1, 2 ..$ is:

$$B_{n,m}^{XT}(p) = p^{m-n}(1-p)^{2n-m} C_{m-n}^n,$$

where $C_{m-n}^n$ is the number of $(m-n)$-combinations of $n$. The finite quantum efficiency of detection of $k$ incident photons is described by the following POVM:

$$B_{n,k}^{QE}(\eta) = \eta^n (1-\eta)^{k-n} C_n^k$$

where $C_n^k$ is the number of $n$-combinations of $k$. By combining the two, a set of POVMs that would predict the outcome of the MPPC with loss, saturation and crosstalk is obtained:

$$\Pi_N = \begin{cases} \sum_{n=[N/2]}^{N} \sum_{k=n}^{\infty} B_{n,k,N} |k\rangle\langle k|, & N < N_{max} \\ I - \sum_{k=0}^{N_{max}} \Pi_k & , N = N_{max} \end{cases} \quad (7)$$

where

$$B_{n,k,N}(\eta, p) = p^{N-n}(1-p)^{2n-N} C_{N-n}^n \eta^n (1-\eta)^{k-n} C_n^k.$$

The $m$-moment operator of the photocount is given by

$$\widehat{N^m} = \sum_{N=0}^{N_{max}} N^m \Pi_N. \quad (8)$$

Eq.(8) allows calculation of a mean and a variance of photocounts, which will be used in Section 5 to reveal suppression of intensity fluctuations of two-mode squeezed vacuum states.

## 3. Assessing photon bunching with MPPC

In this section we describe an experiment, where we use the MPPC for measurements of the second order correlation function (CF) of coherent and squeezed vacuum (SV) states. Let us consider the PDC in frequency degenerate collinear regime. In this case the signal and idler photons are indistinguishable, and the state produced is a *single-mode squeezed vacuum*, with the state vector given by a superposition of even-number Fock states [41]:

$$|\psi\rangle = \sum_{n=0}^{\infty} C_{2n}|2n\rangle \quad (9)$$

The index $n$ is related to the ensembles of modes, accessed via multimode detection [41]. The state vector of a coherent state is

$$|\psi_{Coh}\rangle = \sum_{n=0}^{\infty} C_n|n\rangle \quad (10)$$

The probability amplitudes in Eq.(9, 10) obey the Poissonian distribution, with $|C_n| = \sqrt{e^{-\langle n \rangle}\langle n \rangle^n/n!}$ and $|C_{2n}| = \sqrt{e^{-\langle 2n \rangle}\langle 2n \rangle^{2n}/2n!}$ where $\langle n \rangle$, $\langle 2n \rangle$, are the mean photon numbers, respectively. From Eqs.(1, 9, 10), one can show that for the case of multimode detection, the measured $g^{(2)} = 1 + M/\langle n \rangle$ for the SV state, where $M$ is the inverse number of detected frequency and spatial modes, while $g^{(2)} = 1$ for the coherent state. However, as one can see from Eq.(6), the crosstalk modifies the initial second-order correlation function.

The detailed description of the experimental setup is presented elsewhere [31]. In brief, a ns-pulsed Nd:YAG laser at 266 nm is used to pump a BBO crystal, where PDC occurs, see Fig.2. Collinear PDC photons at 532 nm are directed to a MPPC module (Hamamatsu C10751-02, 400 pixels). The MPPC is sealed into a custom-made housing with a coated optical window and is thermoelectrically cooled down to -4.5°C to reduce the dark noise. A digital oscilloscope is used to capture the analogue output from the MPPC and plot the histogram of its amplitude. The results of the measurements on the coherent state are obtained with a pulsed laser at 532 nm. The MPPC dark noise and ambient luminescence of the setup are measured by extinguishing PDC by tilting the polarization of the pump at 90 degrees by a HWP. The noise is estimated to be 0.001 photon/pulse, and it is subtracted from the PDC signal. The measurement time is increased for the values of photocounts in the range of less than 0.01 photon/pulse in order to mitigate the statistical impact of the noise.

The results of the MPPC measurements are compared with the ones obtained by a traditional HBT experiment, by using a 50/50 beam splitter and two APDs, whose outputs are addressed to a coincidence circuit.

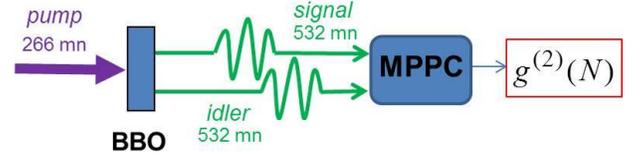

Fig.2 Schematic of the experimental setup for measurement of photon bunching. PDC in collinear regime is generated in a BBO crystal, which is pumped by a pulsed UV-laser (pump). PDC photons (signal and idler) with the same wavelengths impinge onto the MPPC. $g^{(2)}$ is inferred from the MPPC response histogram using Eq.(3).

First, measurements are carried out with the MPPC by measuring coherent states produced by the laser. The dependence of $g^{(2)}$ on the mean number of photocounts, calculated using Eq.(3), is shown in Fig.3a (black dashed trace, squares). We consider the Poissonian statistics of photocounts. The standard deviation of each photon-number component is calculated as the square root of the total number of counts. The standard deviations values are propagated according to Eqs.(3, 6) for calculation of corresponding error bars for the crosstalk probability and $g^{(2)}$. The curve is fitted by Eq.(6) with $p$ being the only fitting parameter (here we consider only linear terms in $p$). The value of the crosstalk probability is found to be $p = 0.177 \pm 0.003$. As shown in Fig.3a, although excess two-photon correlations are not expected for the coherent state ($g_0^{(2)} = 1$), the dependence of CF obtained by the MPPC exhibits strong photon bunching due to the crosstalk.

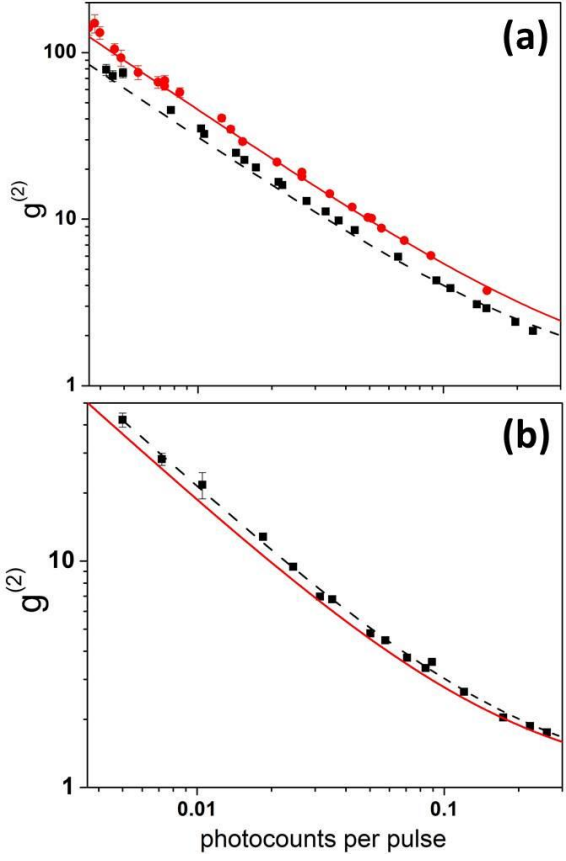

Fig.3 (a) Dependence of $g^{(2)}$ on the mean number of photocounts per pulse, obtained via MPPC for the coherent state (black dashed trace, squares) and for the single-mode SV state (red solid trace, circles). The curves are experimental fits. (b) Comparison of the dependence of the inferred $g^{(2)}$ on the mean number of photocounts per pulse for single-mode squeezed vacuum, measured by the MPPC (red solid trace), with the one obtained with a traditional HBT setup (black dashed trace, squares).

Next, measurements of the single-mode SV state are made. Analogous to the previous case, the dependence of $g^{(2)}$ is calculated from the MPPC amplitude histograms according to Eq.(3). The result is shown in Fig.3a (red solid trace, circles) and fitted by Eq.(6). From the obtained results it is clearly seen that on top of the crosstalk $g^{(2)}$ reveals additional two-photon correlations, which are attributed to the two-photon nature of the SV.

In order to infer the actual $g_0^{(2)}$ of the SV state from the MPPC measurements, Eq.(6) is used with the value of $p$ obtained from the earlier measurements of the coherent state. The resulting curve is shown in Fig.3b (red solid trace). The results are compared with the ones obtained in the HBT experiment, which are also shown in Fig.3b (black dashed trace, squares). The results of the measurements with MPPC and HBT setup show reasonable agreement. Slight deviation is most likely caused by inaccuracies in adjustment of relative quantum efficiencies of MPPC and APDs.

The obtained results demonstrate that with a proper account for the crosstalk, MPPC detectors can be used for measurements of the second-order correlation function. It is worth mentioning that a similar algorithm can be used for the measurement of higher-order correlation functions by using just a single MPPC [31].

## 4. Calibration of MPPC crosstalk using coherent states

As it was demonstrated in the previous section, the crosstalk modifies the initial statistics of photons, and that is why its accurate calibration is needed. A widely implemented approach is based on the measurement of the photocount distribution of dark noise [34, 42]. Assuming that dark counts obey a Poisson distribution, one can calculate the expected mean number of dark counts $\langle N \rangle_{DC}$. The probability of the crosstalk is calculated as the difference between the number of single photon events expected from the Poisson distribution with the mean $\langle N \rangle_{DC}$ and the actual number of single photon events observed for the dark noise. This approach allows calibration of the crosstalk without much effort. However, the method is highly sensitive to statistical errors for the case of low-noise detectors, which are of most practical interest.

Other methods rely on statistical modeling of the measured photon statistics of well characterized light sources and solving an inverse problem. However, in the analysis either the probability of crosstalk has been analyzed as one of several fitting parameters [30] or the assumption has been made that for a given pixel only one crosstalk event could occur, which may not always be the case [28, 31, 43].

Here we develop an alternative approach which is based on the measurements of the second order correlation function. From Eq.(6) it follows that the crosstalk probability can be found from measurements of the correlation function for some source with known statistics, such as, a coherent source with $g_0^{(2)} = 1$. Remarkably, since $g^{(2)}$ is insensitive to optical losses, the method does not require *a-priori* knowledge of the quantum efficiency of the detector under test. Here we experimentally test the approach with several commercially available MPPCs.

We use a ns-pulsed Nd:YAG laser at 532 nm which provides a good quality coherent state ($g_0^{(2)} = 1.010 \pm 0.002$). The laser beam is attenuated by a variable filter and then impinges onto the MPPC, see Fig.4. Three different MPPC's (Hamamatsu, C10507-11-XXXU series) with 100, 400, 1600 pixels per chip with the corresponding pixel sizes of 100x100 µm², 50x50 µm², and 25x25 µm² are studied. The temperature of MPPC can be varied thermoelectrically from the room temperature down to -8ºC. The dark noise is measured by blocking the laser beam and then subtracted from the signal. The signals from MPPCs are digitized and used for evaluation of $g^{(2)}$. The experimental dependence of $g^{(2)}(N_{Total,CT})$ is fitted with Eq.(6) with $p$ being a single fitting parameter.

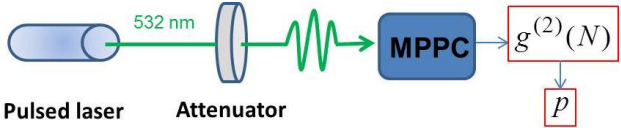

Fig.4 Schematic of the experimental setup for calibration of the MPPC crosstalk probability. A pulsed laser at 532 nm is used as a source of coherent states. Attenuator allows controlling the average number of photons impinging onto MPPC. Three different MPPCs have been tested at different temperatures. Crosstalk probability is calculated from the measured $g^{(2)}$ using Eq.(6).

The resulting dependencies of $g^{(2)}(N_{Total,CT})$ for three different detectors at -4.5ºC are shown in Fig.5. The crosstalk probabilities $p$, are obtained from fits of the experimental data which yield typical values of coefficient of determination (COD) 0.983-0.99. For the sake of future comparison with the conventional method of crosstalk calibration using dark noise $p_{DC}$ [34, 42], we calculate the expression $p + 2p^2$, which is shown in Table 1.

Table 1. Experimental results of crosstalk calibration

| Pixel size, µm² | Dark counts/pulse | $p+2p^2$ | $p_{DC}$ |
|---|---|---|---|
| 100x100 | 0.021 | 0.87±0.01 | 0.610±0.015 |
| 50x50 | 0.008 | 0.21±0.005 | 0.23±0.03 |
| 25x25 | 0.002 | 0.07±0.002 | 0.10±0.11 |

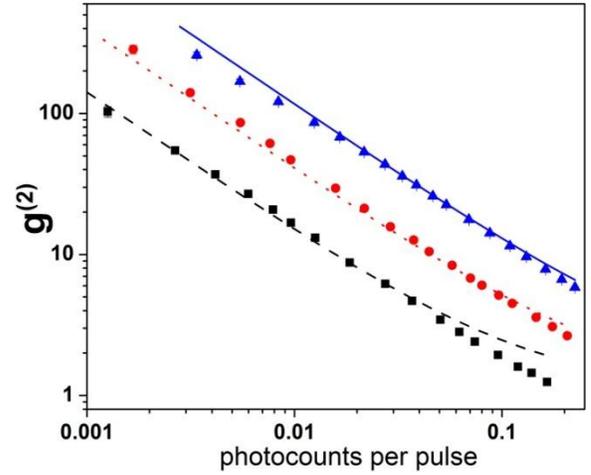

Fig.5 Dependencies of $g^{(2)}$ on the mean number of photocounts per pulse measured by the MPPC at -4.5ºC with 25x25 µm² (black squares, dashed line), 50x50 µm² (red circles, dotted line), and 100x100 µm² pixel size (blue triangles, solid line). Lines are fits to experimental data.

From the presented results it follows, that those detectors with larger pixel size have larger crosstalk. This result is qualitatively confirmed by other groups using alternative calibration techniques, and explained by the quadratic dependence of the crosstalk probability on the gain of the detector [44, 45]. The gain of each pixel is given by $G = C\Delta V$, where $C$ is the capacitance of the pixel and $\Delta V = V_{op} - V_{bd}$, where $V_{bd}$ is the breakdown voltage and $V_{op}$ is the operational voltage [44]. Detectors with larger size of the pixels have larger capacitance, and consequently exhibit larger probability of crosstalk.

Additionally, the crosstalk is measured at four different temperatures of 25ºC, 12.5ºC, 5ºC, - 4.5ºC for the MPPC with 50x50 µm² pixel size. The estimated crosstalk probabilities are obtained similar to the above and shown in Fig.6 (red squares). The dependence of the crosstalk probability on the temperature is again attributed to the corresponding change of the gain [34]. The results are compared with those obtained by the method based on the measurement of dark counts. For each detector under test, the number of dark counts measurements is intentionally chosen to be ~40 % larger than the number of measurements of $g^{(2)}$. Even in this case, the method based on the measurement of $g^{(2)}$ gives the crosstalk probability with significantly smaller experimental uncertainty, see Fig.6. Thus the method represents considerable practical interest for characterization of low-noise MPPCs.

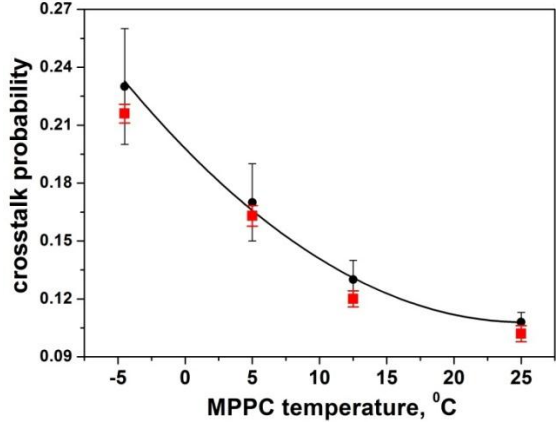

Fig.6 Dependence of crosstalk probability found from $g^{(2)}$ measurements (red squares), and measured from the dark noise (black circles) on temperature for the MPPC with 50x50 μm² pixel size. Solid line is a fit with a quadratic function.

## 5. Measurement of two-mode squeezing with MPPC

So far we considered the use of MPPC in the measurements of squeezed vacuum (SV) states, in the regime of less than one photon per pulse. In this section we describe the application of the MPPC to the study of relatively bright SV states.

Let us consider PDC in frequency non-degenerate regime. The state vector, of the SV state, can be written in the Fock-state basis as follows [46]:

$$|\psi\rangle = \sum_{n=0}^{\infty} C_n |n\rangle_s |n\rangle_i, \quad (11)$$

where $|n\rangle_j$ denotes the Fock-state of $n$ photons in the $j$-th mode, where $j = s, i$ denote signal and idler modes respectively, and $|C_n|^2$ is the probability amplitude. Strong correlation between the photon numbers in the signal and idler modes results in the suppression of the variance of their difference below the classical limit [47, 48]. This effect, referred to as a *two-mode squeezing*, can be quantified by the noise reduction factor (*NRF*), given by

$$NRF = \frac{Var(\widehat{N_s} - \widehat{N_i})}{\langle \widehat{N_s} + \widehat{N_i} \rangle}, \quad (12)$$

where $\widehat{N_s}$ and $\widehat{N_i}$ are the photocounts of the detectors in the signal and idler modes respectively, and $\langle \hat{X} \rangle$ stands for the mean value of an observable $\hat{X}$ for a given quantum state. For the coherent light photon numbers in two modes are statistically independent and thus NRF is equal to unity. In the PDC process the number of photons in conjugated modes are strongly correlated and NRF is less than unity. Thus NRF measurements allow convenient characterization of multiphoton entangled states.

The two-mode squeezing can be studied with various types of photodetectors, depending on the photon fluxes of the PDC. At a moderate gain, when the SV state contains several photons per pulse, it is essential that the detector is able to resolve several simultaneously impinging photons. Here we use the MPPC to directly observe the two-mode squeezing of relatively bright SV state of up to almost 10 photons per pulse.

The variance of the difference of photocount numbers in signal and idler modes is given by

$$Var\langle \widehat{N_s} - \widehat{N_i} \rangle = \langle \widehat{N_s^2} \rangle - \langle \widehat{N_s} \rangle^2 + \langle \widehat{N_i^2} \rangle - \langle \widehat{N_i} \rangle^2 - 2\langle \widehat{N_s} \widehat{N_i} \rangle + 2\langle N_s \rangle \langle N_i \rangle \quad (13)$$

From Eqs.(7, 8, 12, 13) the NRF can be calculated for any given two-mode state. Further we assume that two MPPCs, used in joint detection, have the same values of $\eta$, $p$, and $N_{max}$. When $\langle \hat{n} \rangle \to 0$ the model yields $NRF = \frac{1+3p}{1+p}$ for the coherent state which is greater than unity—the value expected for a lossy PNRD without the crosstalk. Thus, the effect of the crosstalk for small photon numbers results in the increase of the $NRF$. Also, as $\langle \hat{n} \rangle \to 0$ for the SV state the model yields $NRF = \frac{1+3p}{1+p} - (1+p)\eta$. Thus the difference in $NRF$ values for the coherent and SV states is equal to the quantum efficiency $\eta_{eff} \equiv (1+p)\eta$ of the detectors differing from the absolute q.e. due to the presence of crosstalk [32, 49]. Numerical calculations show that the effect of saturation decreases the $NRF$ with increasing $\langle \hat{n} \rangle$ (see Fig.8a).

The experimental setup is shown in Fig.7. The 4th harmonic of Nd:YAG pulsed laser at 266 nm (repetition rate 20kHz, 30 nsec pulse width) is used as a pump two 5 mm length BBO crystals where PDC occurs. Each crystal is cut for collinear frequency non-degenerate type I PDC with signal and idler modes at $\lambda_s = 500$ nm and $\lambda_s = 568$ nm, respectively. The half-wave plate and polarization beamsplitter (PBS) (not shown) are used to change the pump power. After passing through the BBOs the pump beam is rejected by a UV-mirror while

PDC radiation passed through it. The signal and idler beams are split by a dichroic mirror (DM): the former is transmitted and the latter is reflected. At each arm there are lenses ($f$=500 mm) with their front focuses at the BBOs. Two iris apertures are set according to $d_s/d_i = \lambda_s/\lambda_i$ after the lenses, selecting the conjugated spatial modes [48]. The irises are set to 10 mm and 11.36 mm in the arms with 500 nm and 568 nm, respectively. The spatially selected spectral bandwidth of PDC is calculated to be 3 nm. The PDC is then collected by lenses ($f$=200 mm) and formed a beam spot of 1.2 mm at the MPPC chip. The ambient light is suppressed by means of interference filters centered at 500 and 568 nm, respectively, both being 15 nm FWHM and a peak transmission of 95%. Two identical MPPCs, model Hamamatsu 10751-02, are used. The saturation behavior is checked by measuring the dependence of photocounts versus the number of incident photons (see inset in Fig.8a), and it is found to be identical for both MPPCs. The crosstalk values are checked by the method described in Section 4, and the values are found to be identical. A 0.1dB neutral density filter is introduced in one of the arms to equilibrate the quantum efficiencies of MPPCs at different wavelengths. Both MPPCs are cooled to reduce the dark noise, see Section 3. The signals from the MPPCs are digitized by an AD card (NI PCI-5154) and discriminated according to their amplitudes. The AD card is synchronized with a pump laser and detection window was set to 70 ns. For each power level $10^6$ triggers are collected. The data analysis is done using LabView and Mathlab software. The MPPC dark noise and ambient luminescence of the setup are measured similarly to the procedure described in Section 3. The measurements performed at different pump powers provide the dependence of NRF and $g^{(2)}$ on the number of photocounts per pulse.

Similar measurements are performed for the coherent light emitted from an attenuated cw Nd:YAG laser ( at 532 nm) which is chopped by an acoustic optical modulator (AOM) at the frequency of 20 kHz with the pulse duration of 30 ns. Then it is fed into a single mode fiber and collimated by a lens (f=6,24 mm) at the MPPCs. Interference filters centered at 532 nm with 1.5 nm FWHM are used to filter out an ambient light.

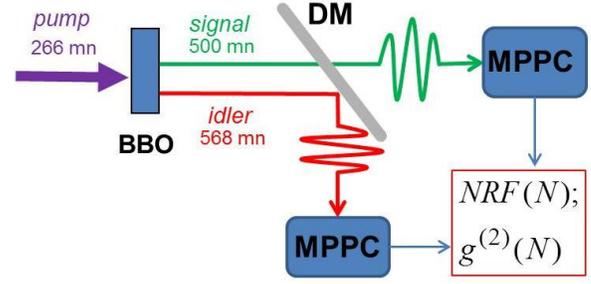

Fig.7 Schematic of the experimental setup for measurement of the two-mode squeezing. PDC in collinear regime is generated in a BBO crystal, which is pumped by a pulsed UV-laser (pump). PDC photons (signal and idler) have different wavelengths. They are split by a dichroic mirror (DM) and addressed to two MPPCs. $NRF$ and $g^{(2)}$ are calculated from joint statistics of photocounts.

Fig.8a shows the experimental data and theoretical fits of $NRF$ versus $\langle \hat{n} \rangle$ for the coherent and SV states. From the difference of the measured $NRFs$ for the two states close to $\langle \hat{n} \rangle = 0$, one finds the experimental value of the effective quantum efficiency for the SV state at 568 nm. The value yields $\eta_{eff,SQZ}^{exp} = 0.17 \pm 0.03$, and it isfurther used for the conversion of photocounts to photon numbers. Effective quantum efficiency for the coherent state at 532 nm is found from $\eta_{eff,SQZ}^{exp}$ assuming the dependence of quantum efficiency on the wavelength, provided by the manufacturer, and yielding $\eta_{eff,COH}^{exp} = 0.20 \pm 0.03$. The theoretical fits of the data to the model yield $N_{\max} = 3, p = 0.28 \pm 0.005$, and $\eta = 0.163 \pm 0.007$ for the coherent state, and $N_{\max} = 3, p = 0.30 \pm 0.02$ and $\eta = 0.145 \pm 0.009$ for the SV state. These parameters yield $\eta_{eff,COH}^{fit} = 0.21 \pm 0.01$ for the coherent state, and $\eta_{eff,SQZ}^{fit} = 0.19 \pm 0.01$, which are in good agreement with the experimental values of $\eta_{eff}$ used for the calculation of photon numbers.

The results clearly demonstrate that $NRF$ gradually decreases with the increasing $\langle \hat{n} \rangle$ for both studied states, which is caused by the saturation of the MPPC. At the same time, the $NRF$ for the SV state lies below the one for the coherent state—a signature of squeezing. The crosstalk elevates the $NRF$ at low $\langle \hat{n} \rangle$ to values above unity, as predicted by the theory. Surprisingly, the obtained value of $N_{max}$ appears to be much smaller than the one expected from the MPPC with 400 pixels. We have found that the MPPC modules do not allow reliable discrimination of photon numbers beyond $\langle \hat{n} \rangle > 5$ due to the limitations of the built-in electronic circuitry, which is inaccessible for adjustment. This might also be the cause of the deviation of the measured NRF from the theoretical prediction for $\langle \hat{n} \rangle > 5$. Note that reliable photon number resolution up to 20 photons was demonstrated with similar detectors using customized electronics [50]. Also, the obtained value of $p$ is higher than an earlier result, obtained for the same detector, see Section 4, [32] which can be due to the presence of higher-order crosstalk events which are not considered in the present model.

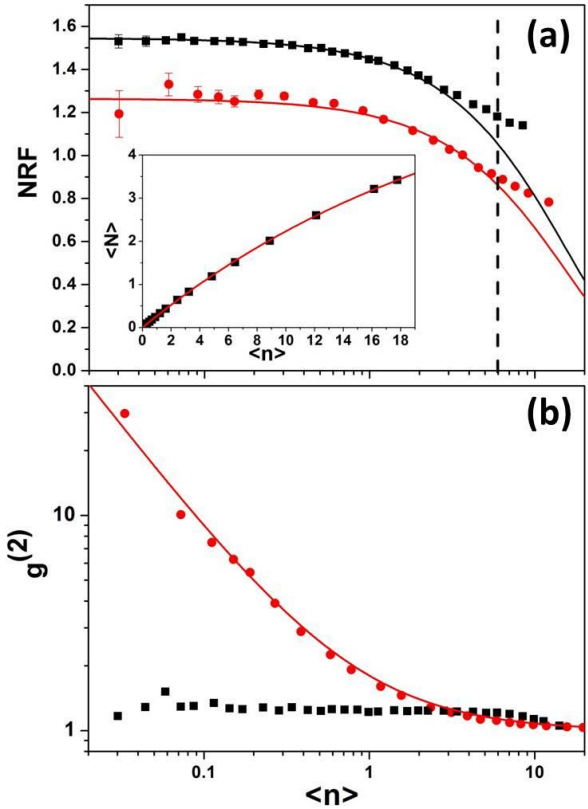

Fig.8 (a) Dependence of the NRF on the mean number of photons measured for the coherent state (black squares) and the SV state (red circles). The theoretical fits for the NRF are performed in the range of up to 6 photons (vertical dashed line), which marks the range of reliable photon number resolution. The inset shows the mean number of photocounts versus the mean number of incident photons in one of the MPPCs for the coherent state. (b) Dependence of $g^{(2)}$ on the mean number of photons measured for coherent light (black squares) and squeezed vacuum (red circles).

It is also of interest to analyze the dependence of the second order intensity correlation function, defined here as $g^{(2)} = <N_s N_i>/<N_s><N_i>$ [37]. For PDC it is shown in Fig.8b (red circles), and demonstrates the rise with the decreasing of the number of photons per pulse, according to $g^{(2)} = 1 + M/\langle n \rangle$, where $M = 0.808$ is an inverse number of detected modes. At the same time, $g^{(2)}$ for coherent states was almost constant, yielding $g^{(2)} = 1.20 \pm 0.09$, despite of the cross-talk and the dark noise which still presents in both MPPCs (black squares). It is explained by the fact that the crosstalk and dark counts in two MPPCs are uncorrelated and do not contribute to $g^{(2)}$. It is worth to compare this result with the one obtained in [31, 32], where only a single MPPC was used for the measurements of $g^{(2)}$. In that case the crosstalk has added considerable "false" correlations, mimicking in this way the true photon bunching.

## Conclusions

In conclusion, we experimentally studied detection of various non-classical states of light with photon number resolving MPPCs. Based on the developed algorithm of obtaining the second order correlation function $g^{(2)}$ and the nonlinear crosstalk model, we measured $g^{(2)}$ for coherent and squeezed vacuum states. It was demonstrated that the crosstalk leads to additional terms in the correlation function for both of the studied states. The presented analysis can be directly applied to other types of spatially resolving detectors such as, single photon resolving ICCD and EMCCD, as well as for measurements of higher order correlation functions.

We also presented a new accessible method for calibration of the MPPC crosstalk probability. The method relies on the fundamental properties of coherent states, and the generalized crosstalk model. It was tested for three commercial MPPCs with different pixel sizes and at different temperatures. The new method exhibited much less uncertainty in determination of the crosstalk probability compared to conventional method based on measurements of dark noise. The method may represent practical interest for applications, and future development of optimized low noise MPPCs.

Finally, we used MPPCs to reveal the two-mode squeezing of relatively bright SV states. We developed a theoretical POVM model, which accounts for MPPC saturation and finite quantum efficiency. We analyzed fluctuations of the difference of the intensities in two modes by measuring NRFs for coherent and SV states.

We demonstrated that the crosstalk in the MPPC leads to an increase of the *NRF* for both coherent and SV states. Moreover we analyzed the effect of the MPPC saturation, which leads to the decrease of *NRF* with increasing photon numbers. We also showed that, in contrast to the measurements by a single MPPC, $g^{(2)}$ was not affected by the crosstalk. The experimental data agrees with the trend of the presented theoretical model which takes into account the saturation and the crosstalk. Our results extended the use of MPPCs for the characterization of quantum light in a significantly broader dynamic range than that of conventional SPADs, thus allowing access to affordable detectors with photon number resolution.

Currently, implementation of the MPPC technology is restricted by several technical issues, where the problem of the crosstalk is the major one. Development of advanced techniques of chip fabrication and better optical isolation of pixels, will allow to mitigate the problem of crosstalk, and open the way towards implementation of the MPPC technology in the broad range of applications ranging from the studies of atomic-photon interfaces to astronomy applications.

## Acknowledgements

We would like to thank Maria V. Chekhova, Timur Sh. Iskhakov and Si Hui Tan for their contributions at various stages of this project. We acknowledge financial support by A-STAR Investigatorship grant.

## References and links

1. E. Knill, R. Laflamme, G.J. Milburn, "A scheme for efficient quantum computation with linear optics," Nature **409**, 46–52 (2001).
2. P. Kok, W.J. Munro, K. Nemoto, T.C. Ralph, J.P. Dowling, G.J. Milburn, "Linear optical quantum computing with photonic qubits," Rev. Mod. Phys. **79**, 135–175 (2007).
3. J.L. O'Brien, "Optical quantum computing," Science **318**, 1567–1570 (2007).
4. M.W. Mitchell, J.S. Lundeen, A.M. Steinberg, "Super-resolving phase measurements with a multiphoton entangled state," Nature, **429**, 161-164 (2004).
5. P. Walther, J.-W. Pan, M. Aspelmeyer, R Ursin, S. Gasparoni, A. Zeilinger, "De Broglie wavelength of a non-local four-photon state," Nature, **429**, 158-161 (2004).
6. I.B. Bobrov. D.A. Kalashnikov, and L.A. Krivitsky, "Imaging of spatial correlations of two-photon states", Phys. Rev. A **89**, 043814 (2014).
7. L. Pezzé and A. Smerzi, "Mach-Zehnder interferometry at the Heisenberg limit with coherent and squeezed vacuum light," Phys. Rev. Lett. **100**(7), 073601 (2008).
8. G. Brassard, N. Lütkenhaus, T. Mor, and B.C. Sanders "Limitations on practical quantum cryptography," Phys. Rev. Lett. **85,** 1330–1333 (2000).
9. D. Bouwmeester, J.-W. Pan, K. Mattle, M. Eibl, H. Weinfurter, and A. Zeilinger "Experimental quantum teleportation," Nature **390,** 575–579 (1997).
10. M. Żukowski, A. Zeilinger, M.A. Horne, and M.A. Ekert, "Event-ready-detectors' Bell experiment via entanglement swapping," Phys. Rev. Lett. **71,** 4287–4290 (1993).
11. F. Dell'Anno, S. De Siena, F. Illuminati, "Multiphoton quantum optics and quantum state engineering," Physics Reports **428**, 53-168 (2006).
12. J.-W. Pan, Z.-B. Chen, C.-Y. Lu, H. Weinfurter, A. Zeilinger, and M. Zukowski, "Multiphoton entanglement and interferometry," Rev. Mod. Phys. **84**, 777-838 (2012).
13. R. H. Hadfield, "Single-photon detectors for optical quantum information applications" Nature Photonics, **3**, 696-705, (2009).
14. D. Achilles, C. Silberhorn, C. Sliwa, K. Banaszek, and I.A. Walmsley, "Fiber assisted detection with photon-number resolution," Opt. Lett. **28**, 2387–2389 (2003).
15. M. J. Fitch, B. C. Jacobs, T. B. Pittman, J. D. Franson, "Photon-number resolution using time-multiplexed single-photon detectors," Phys. Rev. A **68**, 043814 (2003).
16. M. Mičuda, O. Haderka, M. Ježek, "High-efficiency photon-number-resolving multichannel detector," Phys Rev A **78**, 025804 (2008).
17. S. Takeuchi, J. Kim, Y. Yamamoto, H.H. Hogue, "Development of a high quantum-efficiency single-photon counting system," Appl. Phys. Lett. **74**, 1063–1065 (1999).
18. J. Kim, S. Takeuchi, Y. Yamamoto, H.H. Hogue, "Multiphoton detection using visible light photon counter," Appl. Phys. Lett. **74**, 902–904 (1999).
19. E. Waks, K. Inoue, E. Diamanti, Y. Yamamoto, "High-efficiency photon-number detection for quantum information processing," IEEE J. Sel. Top. Quant. **9**, 1502–1511 (2003).
20. B. Cabrera, R. M. Clarke, P. Colling, A. J. Miller, S. Nam, R. W. Romani, "Detection of single infrared, optical, and ultraviolet photons using superconducting transition edge sensors," Appl. Phys. Lett. **73**, 735 (1998).
21. D. Rosemberg, A.E. Lita, A.J. Miller, S.W. Nam, "Noise-free high-efficiency photon-number-resolving detectors," Phys Rev A, **71**, 061803R (2005).
22. A.E. Lita, A.J. Miller, S.W. Nam, "Counting near-infrared single- photons with 95% efficiency," Opt. Express **16**, 3032-3040 (2008).
23. D. Sahin, and *et al.* "Waveguide photon-number-resolving detectors for quantum photonic integrated circuits," Appl. Phys. Lett, **103**, 111116 (2013).
24. M. Bondani, A. Allevi, A. Agliati, A. Andreoni, "Self-consistent characterization of light statistics," J. Mod. Opt **56**, 226 - 231, (2009).
25. B. E. Kardynal, Z. L. Yuan A. J. Shields, "An avalanche-photodiode-based photon-number-resolving detector," Nature Photonics **2**, 425-428 (2008).

26. O. Thomas, Z.L. Yuan and A.J. Shields "Practical photon number detection with electric field-modulated silicon avalanche photodiodes," Nature Communication **3**, 644 (2012).
27. Hamamatsu web-page http://jp.hamamatsu.com/
28. I. Afek, A. Natan, O. Ambar, Y. Silberberg, "Quantum state measurements using multipixel photon detectors," Phys. Rev. A **79**, 043830 (2009).
29. S.-H. Tan, L. A. Krivitsky, B. –G. Englert "Measuring Quantum Correlations using Lossy Photon-Number-Resolving Detectors with Saturation" Pre-print: arXiv:1210.8022.
30. M. Ramilli, A. Allevi, V. Chmill, M. Bondani, M. Caccia, A. Andreoni, "Photon-number statistics with silicon photomultipliers," J. Opt. Soc. Am. B **27**, 852-862 (2010).
31. D.A. Kalashnikov, S.-H. Tan, M.V. Chekhova, L.A. Krivitsky, "Accessing photon bunching with photon number resolving multi-pixel detector," Opt. Express **19**, 9352-9363 (2011).
32. D.A. Kalashnikov, S.-H. Tan, L.A. Krivitsky, "Crosstalk calibration of multi-pixel photon counters using coherent states", Opt. Express **20**, 5044-5051 (2012).
33. D. A. Kalashnikov, S.-H. Tan, T. Sh. Iskhakov, M. V. Chekhova, and L. A. Krivitsky "Measurement of two-mode squeezing with photon number resolving multipixel detectors", Opt. Lett. **37**, 2829-2831 (2012).
34. A. Vacheret and et al. "Characterization and Simulation of the Response of Multi Pixel Photon Counters to Low Light Levels," Nuclear Instruments and Methods in Physics Research A **656**, 69-83 (2011).
35. A. Spinelli and A. L. Lacaita, "Physics and numerical simulation of single photon avalanche diodes," IEEE Trans. Electron. Dev. **44**(11), 1931–1943 (1997) .
36. I. Rech, A. Ingargiola, R. Spinelli, I. Labanca, S. Marangoni, M. Ghioni, and S. Cova, "Optical crosstalk in single photon avalanche diode arrays: a new complete model," Opt. Express **16**, 8381–8394 (2008).
37. R.J. Glauber, *Quantum Optics and Electronics* (Gordon and Breach, New York, 1965).
38. M. Avenhaus, K. Laiho, M. V. Chekhova, and C. Silberhorn, "Accessing higher order correlations in quantum optical states by time multiplexing," Phys. Rev. Lett. **104**, 063602 (2010).
39. O. Haderka, J. Perina, Jr., M. Hamar, and J. Perina, "Direct measurement and reconstruction of nonclassical features of twin beams generated in spontaneous parametric down-conversion," Phys. Rev. A **71**(3), 033815 (2005).
40. J.-L. Blanchet, F. Devaux, L. Furfaro, and E. Lantz, "Measurement of sub-shot-noise correlations of spatial fluctuations in the photon-counting regime," Phys. Rev. Lett. **101**, 233604 (2008).
41. D. N. Klyshko, *Photons and Nonlinear Optics* (Gordon and Breach, New York, 1988).
42. M. Yokoyama, A. Minamino, S. Gomi, K. Ieki, N. Nagai, T. Nakaya, K. Nitta, D. Orme, M. Otani, T. Murakami, T. Nakadaira, and M. Tanaka, "Performance of multi-pixel photon counters for the T2K near detectors," Nuclear Instruments and Methods in Physics Research A **622**, 567-573 (2010).
43. P. Eraerds, M. Legré, A. Rochas, H. Zbinden, and N. Gisin, "SiPM for fast photon-counting and multiphoton detection," Opt. Express **15**, 14539–14549 (2007).
44. Hamamatsu web-page *http://jp.hamamatsu.com/products/sensor-ssd/4010/index_en.html*
45. A. L. Lacaita, F. Zappa, S. Bigliardi, and M. Manfredi, "On the bremsstrahlung origin of hot-carrier-induced photons in silicon devices," IEEE Trans. Electron. Dev. **40**(3), 577–582 (1993).
46. D.N. Klyshko, "The nonclassical light", Phys. Usp. **39**, 573-596 (1996).
47. A. Heidmann, R.J. Horowicz, S. Reynaud, E. Giacobino, C. Fabre and G. Camy, "Observation of quantum noise reduction on twin laser beams", Phys. Rev. Lett. **59**, 2555-2557 (1987).
48. I.N. Agafonov, M.V. Chekhova, and G. Leuchs, "Two-color bright squeezed vacuum", Phys. Rev. A **82**, 011801 (2010).
49. I. N. Agafonov, M.V. Chekhova, T. S. Iskhakov, A. N. Penin, G. O. Rytikov, O. A. Shcherbina, "Absolute calibration of photodetectors: photocurrent multiplication versus photocurrent subtraction" Opt. Lett. **36**, 1329 (2011).
50. L. Dovrat, M. Bakstein, D. Istrati, A. Shaham, H. S. Eisenberg, "Measurements of the dependence of the photon-number distribution on the number of modes in parametric down-conversion", Opt. Express **20**, 2266-2276 (2012).